\documentclass[prd,aps,notitlepage,twocolumn,longbibliography,superscriptaddress]{revtex4-1}
\usepackage[dvipsnames, svgnames]{xcolor}
\usepackage{amsfonts,amssymb,amsmath,mathrsfs,graphicx,bm,orcidlink,url,here,mathtools,chngcntr,ulem}
\usepackage{hyperref}

\newcommand{\secl}[1]{\noindent \textbf{#1—}}
\renewcommand{\Im}{\mathrm{Im}}

\begin{document}

\title{Pole Skipping, Avoided Crossing, and Resonant Excitation in\\ Kerr Quasinormal Modes near Algebraically Special Frequencies}

\author{Kei-ichiro Kubota\orcidlink{0000-0002-1576-4332}}
    \affiliation{Institute for Cosmic Ray Research, The University of Tokyo, 5-1-5 Kashiwanoha, Kashiwa, Chiba 277-8582, Japan}
\author{Hayato Motohashi\orcidlink{0000-0002-4330-7024}}
    \affiliation{Department of Physics, Tokyo Metropolitan University, 1-1 Minami-Osawa, Hachioji, Tokyo 192-0397, Japan}
    \affiliation{Yukawa Institute for Theoretical Physics, Kyoto University, 606-8502, Kyoto, Japan}
\date{\today}

\begin{abstract}
Kerr quasinormal modes near algebraically special frequencies exhibit anomalous behavior, including apparent bifurcation, disappearance, and a nonsmooth connection to the Schwarzschild limit, which has remained puzzling for decades. Tracking poles and zeros of Green-function building blocks across different Riemann sheets, we show that the bifurcation is due to an avoided crossing accompanied by resonant excitation, while the disappearance is due to pole skipping caused by cancellation of a quasinormal-mode pole by a Matsubara-mode zero. This resolves the physical origin of these long-standing anomalies.
\end{abstract}

\maketitle

\secl{Introduction}
The ringdown of a perturbed black hole provides one of the cleanest probes of strong-field gravity. 
In linear perturbation theory in general relativity, the ringdown signal is primarily described by a superposition of damped oscillations at characteristic complex frequencies, the quasinormal modes (QNMs).
These frequencies correspond to poles of the Green's function and depend solely on the mass and spin of the underlying Kerr black hole~\cite{Chandrasekhar:1975zza,Leaver:1985ax}. 
Their measurement underlies black-hole spectroscopy and, more broadly, the use of gravitational waves to test the Kerr geometry in the dynamical regime~\cite{Berti:2009kk,Berti:2025hly}. 
This program relies on a precise understanding of the analytic structure underlying black-hole perturbations, including not only the QNM spectrum itself but also the nontrivial spectral structures needed for a consistent physical interpretation of black-hole ringdown.

Among these structures, the algebraically special (AS) frequencies~\cite{Couch:1973zc,Wald:1973wwa,Chandrasekhar:1984mgh} are particularly distinctive: they form a set of purely imaginary frequencies that occur only for gravitational perturbations~\cite{Berti:2009kk,Berti:2025hly}.
Physically, they are related to total-transmission modes (TTMs)~\cite{Andersson:1994tt}, in which the black-hole potential becomes effectively transparent to the corresponding wave.
Because they are specific to gravitational perturbations, the phenomena associated with AS frequencies—including TTMs and the QNMs near them—probe a genuinely gravitational aspect of black-hole scattering and have therefore remained a subject of sustained interest~\cite{Onozawa:1996ux,MaassenvandenBrink:2000iwh,Berti:2004md,Keshet:2007be,Cook:2014cta,Cook:2016fge,Cook:2016ngj,Cook:2018ses,Cook:2022kbb,Tuncer:2025dnp,Yu:2026rku}.

Yet Kerr QNMs near AS frequencies exhibit a long-standing anomaly: the spectrum appears to bifurcate, modes seem to disappear, and the connection to the nonrotating Schwarzschild limit looks nonsmooth.
The best-known example is the $(l,m)=(2,2)$ eighth overtone of the gravitational QNM spectrum. 
Its anomalous behavior can already be traced back four decades to Leaver's original Kerr-QNM computations~\cite{Leaver:1985ax} and has remained a subject of discussion in subsequent studies~\cite{Onozawa:1996ux,MaassenvandenBrink:2000iwh}. 
For positive spin $a>0$, high-precision calculations found two branches, conventionally denoted by $8_0$ and $8_1$, which approach the Schwarzschild eighth overtone as $a\to +0$ but do not converge to it exactly and instead appear to terminate at finite spin~\cite{Cook:2014cta}. 
By contrast, for $a<0$, only a single branch approaches the Schwarzschild overtone. 
This suggested an apparent discontinuity between the Kerr and Schwarzschild spectra near the AS frequency. 
By invoking unconventional Riemann sheets~\cite{Leung:2003eq,Fiziev:2010yy,Fiziev:2011mm}, more recent work has clarified the global structure of the Kerr QNM spectrum~\cite{Chen:2025sbz}.
Here, the ``conventional'' sheet is reached by analytic continuation from the real-frequency axis without crossing the branch cut of the radial Teukolsky solutions, and it hosts the standard QNM spectrum. In contrast, the ``unconventional'' sheet lies beyond this branch cut and is not directly accessible from the physical axis.
However, the physical origin of the near-AS QNM anomaly has remained unclear.
In particular, the global continuation of the spectrum does not by itself explain why a mode becomes invisible in the conventional QNM spectrum,
what triggers the apparent bifurcation, or how these spectral features manifest in the mode excitation near the AS frequency.

In this Letter, we revisit the problem from the perspective of the Green's function and track not only QNM poles but also Matsubara-mode (MM) zeros of Green-function building blocks across conventional and unconventional Riemann sheets. 
Although black-hole spectroscopy has traditionally focused on QNM poles, such zeros have received much less attention; we show that they are essential for resolving the near-AS anomaly. 
The MM frequencies are not merely auxiliary markers: they are the discrete thermal frequencies familiar from finite-temperature field theory~\cite{Matsubara:1955ws,LeBellac1996,Kapusta:2006pm}. 
In black-hole scattering, the same thermal frequency scale is set by the Hawking temperature and horizon chemical potential, thereby linking horizon thermality to ringdown scattering.
They arise not only in the extremal Kerr/CFT context~\cite{Chen:2010ni}, but also directly from the Teukolsky equation for generic subextremal Kerr black holes~\cite{Motohashi:2026mbn}.

This pole-zero perspective reveals a rich structure near the AS frequencies.
We show that the apparent disappearance of the Kerr mode approaching the AS frequency is due to ``pole skipping''~\cite{Grozdanov:2017ajz,Blake:2017ris,Blake:2018leo,Grozdanov:2018kkt} caused by an MM zero, a mechanism recently emphasized in hydrodynamics, holography, and quantum chaos as a structural feature of thermal Green's functions~\cite{Grozdanov:2019uhi,Blake:2019otz,Natsuume:2019sfp,Haehl:2018izb,Natsuume:2020snz,Wang:2022mcq,Natsuume:2019xcy,Natsuume:2021fhn,Ahn:2020baf,Blake:2021hjj,Liu:2020yaf,Abbasi:2019rhy,Abbasi:2020ykq,Abbasi:2020xli,Amano:2022mlu,Jansen:2020hfd,Ahn:2020bks,Ahn:2019rnq,Grozdanov:2020koi,Yuan:2023tft,Grozdanov:2023txs,Grozdanov:2024wgo}. 
We further show that the apparent bifurcation is the visible manifestation of an avoided crossing with another mode on a different Riemann sheet, and that this interaction is accompanied by resonant excitation~\cite{Motohashi:2024fwt}. 

This perspective also connects the problem with the broader language of non-Hermitian spectral theory and scattering resonances. 
In scattering theory, resonances are poles of analytically continued Green's functions or S-matrices, and their trajectories can move between Riemann sheets as a control parameter is varied~\cite{Kukulin1989}. 
More broadly, non-Hermitian physics provides a natural framework for complex spectra and exceptional-mode interactions in open systems~\cite{moiseyev_2011,Ashida:2020dkc}. 
Here, we identify a concrete black-hole realization of this pole dynamics, enriched by Matsubara zeros that can hide QNM poles through pole skipping.

We focus below on gravitational perturbations of the $(l,m)=(2,2)$ multipole as the clearest example.
The key point is not the eighth overtone itself, but the mechanism it exposes: QNM anomalies near AS frequencies can arise from pole-zero dynamics across Riemann sheets rather than from pole trajectories alone.
The corresponding higher-multipole results are presented at the end of the Letter and in Supplemental Material. 
We use units with $c=G=1$ throughout.

\begin{figure}
    \centering
    \href{https://zenodo.org/records/19688891/preview/Inverse_of_the_incidence_coefficient_of_the_Teukolsky_eq_on_Riemann_sheet.mp4}{\includegraphics[width=1.0\columnwidth]{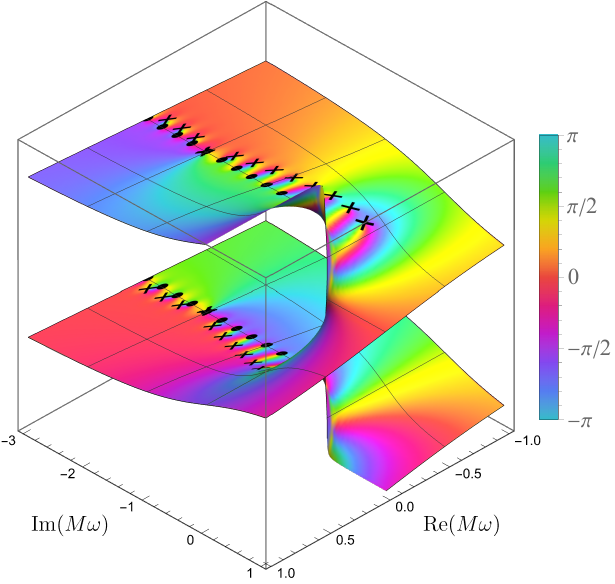}}
    \caption{\label{fig:OneOverAinRiemann}Riemann-surface structure of the Schwarzschild Green's function, visualized by the color-coded phase $\arg(1/A^\mathrm{T}_\mathrm{in})$. 
    Black crosses and dots mark QNM poles and Matsubara zeros, respectively. 
    The prograde and retrograde QNMs lie on the lower and upper sheets, respectively.}
\end{figure}

\begin{figure}[ht!]
    \href{https://zenodo.org/records/19688891/preview/Kerr_black_hole_QNM_frequency_s-2l2m2.mp4}{\includegraphics[width=1.0\columnwidth]{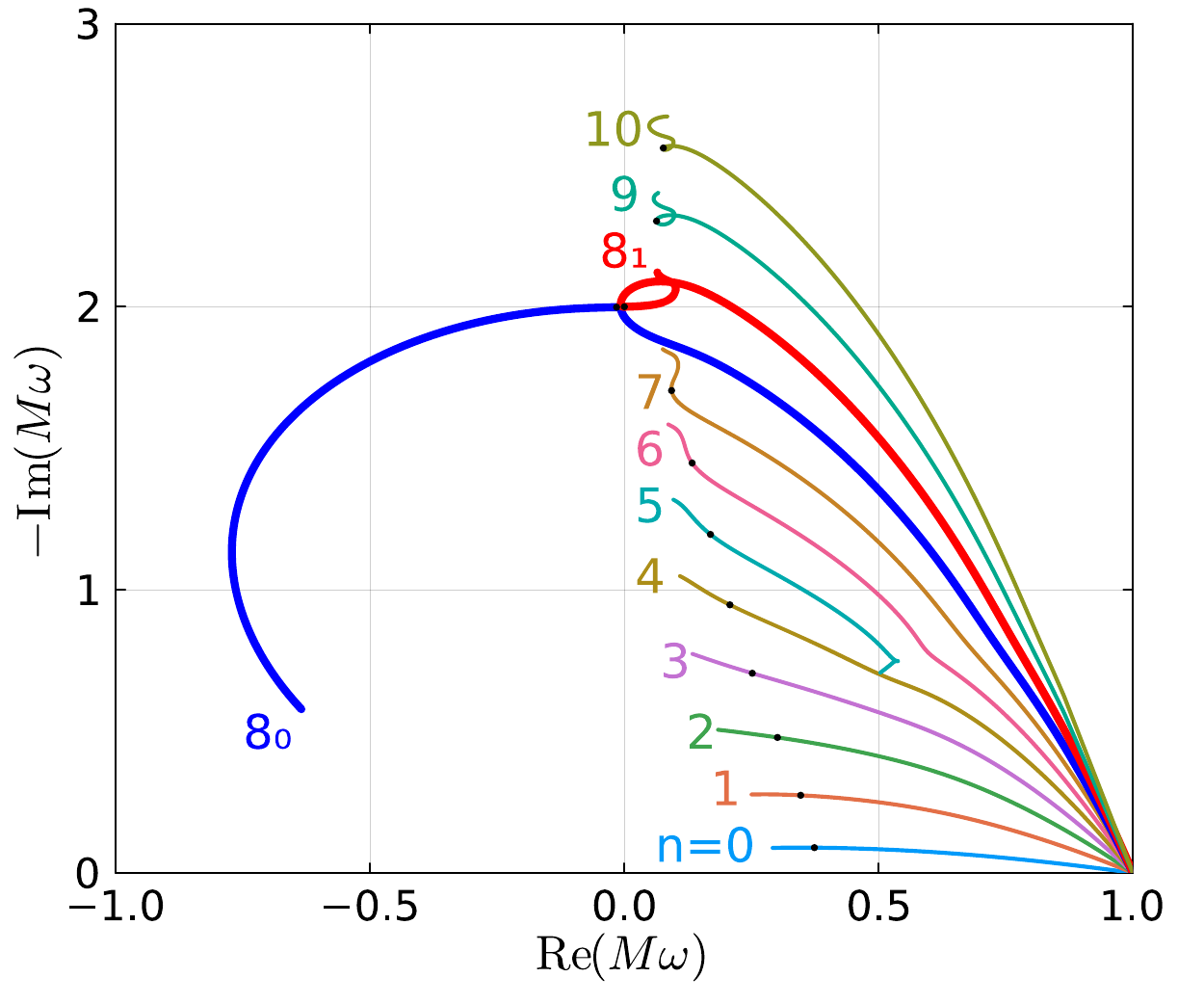}}
    \caption{\label{fig:QNM_l2_m2_frequencies}Prograde Kerr QNM frequencies for $(s,l,m)=(-2,2,2)$ over $-1<a/M<1$. The black dot marks the Schwarzschild limit $a/M=0$.}
\end{figure}

\secl{Apparent Bifurcation, Avoided Crossing, and Resonant Excitation}
In the standard criterion, QNM frequencies are identified with the zeros of $A^\mathrm{T}_\mathrm{in}$, the asymptotic incident amplitude of the in-mode solution for the homogeneous radial Teukolsky equation; the corresponding Green-function contribution is inversely proportional to this amplitude.
The reciprocal amplitude $1/A^\mathrm{T}_\mathrm{in}$ therefore captures the pole structure associated with this criterion.
Figure~\ref{fig:OneOverAinRiemann} illustrates this construction for the Schwarzschild spacetime, showing the phase of $1/A^\mathrm{T}_\mathrm{in}$ mapped onto the resulting Riemann surface.
To access the relevant Riemann sheets, we rotate the branch cut using the Mano--Suzuki--Takasugi (MST) representation~\cite{Mano:1996vt} of $A^\mathrm{T}_\mathrm{in}$~\footnote{Technical details, Kerr snapshots, and animations are given in Supplemental Material A and Ref.~\cite{kubota_2026_18534214}}.
This approach allows us to clarify the analytic structure of not only poles but also zeros of the reciprocal incident amplitude near AS frequencies, including across the branch cut.

For a Schwarzschild black hole of mass $M$, the AS frequencies are~\cite{Couch:1973zc,Chandrasekhar:1984mgh,MaassenvandenBrink:2000iwh}
\begin{align}
    M\omega_\mathrm{AS}^\pm = \pm \mathrm{i} (l-1)l(l+1)(l+2)/12,
    \label{eq:ASfrequency}
\end{align}
where the superscripts $\pm$ correspond to the two signs on the right-hand side.
Mathematically, the Chandrasekhar transformation connecting the Regge-Wheeler (RW) and Zerilli equations becomes singular at these frequencies, leading to a breakdown of isospectrality~\cite{MaassenvandenBrink:2000iwh}.
For the quadrupole case $l=2$, the negative AS frequency is located at $M\omega=-2\mathrm{i}$.

\begin{figure*}[t]
    \begin{minipage}[]{\columnwidth}
        \href{https://zenodo.org/records/19688891/preview/Kerr_black_hole_QNM_frequency_s-2l2m2_zoomedin.mp4}{ \includegraphics[width=1.0\columnwidth]{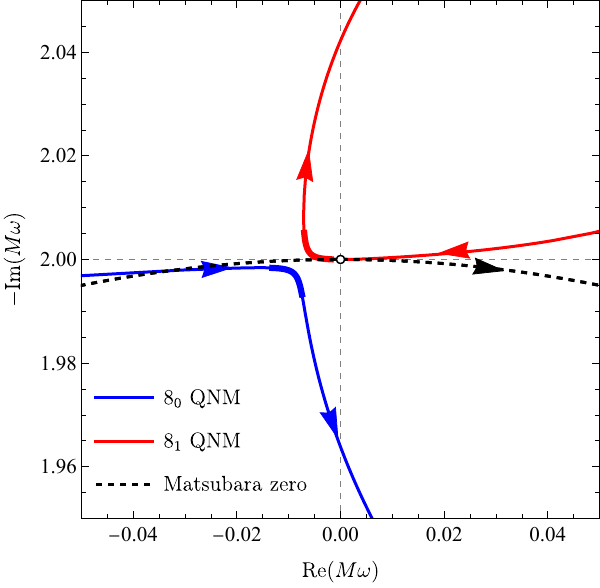}}
    \end{minipage}
    \begin{minipage}[]{\columnwidth}
        \includegraphics[width=1.0\columnwidth]{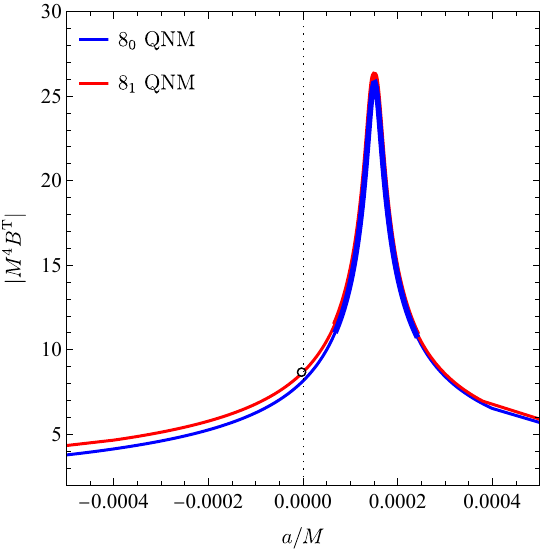}
    \end{minipage}
    \begin{minipage}[]{1.0\textwidth}
        \href{https://zenodo.org/records/19688891/preview/Kerr_black_hole_QNM_excitation_factor_s-2l2m2.mp4}{\includegraphics[width=0.8\textwidth]{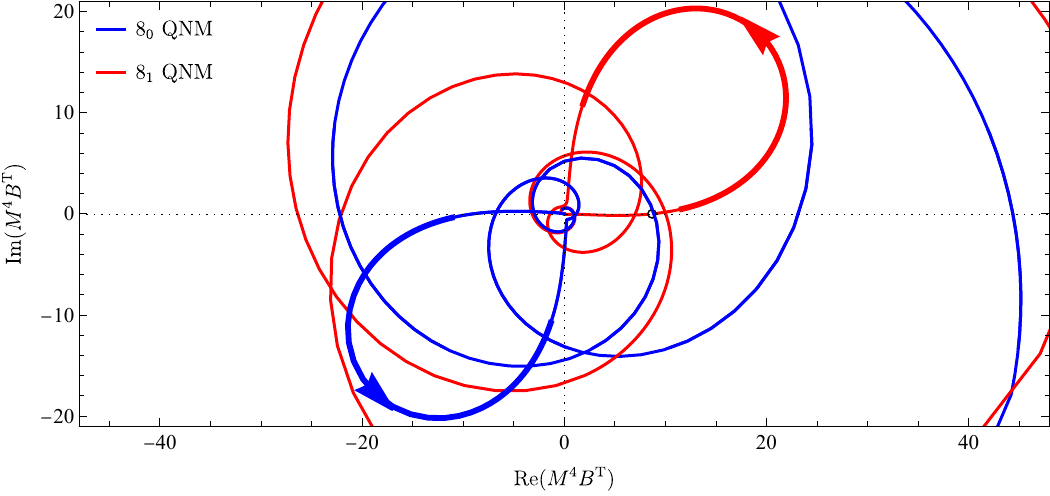}}
    \end{minipage}
    \caption{\label{fig:qnm_and_B}Near-AS QNM and MM frequencies (upper left), absolute values of the excitation factors (upper right), and excitation-factor trajectories in the complex plane (lower) for the $8_0$ and $8_1$ branches. 
    Solid curves denote QNM poles and dashed curves denote Matsubara zeros. 
    The white circle marks the pole-skipping point, and thick curves highlight the resonant-excitation region $7.0 \times 10^{-5} \lesssim a/M \lesssim 2.4 \times 10^{-4}$.}
\end{figure*}

Figure~\ref{fig:QNM_l2_m2_frequencies} shows the trajectory of the QNM poles on the lower sheet of Fig.~\ref{fig:OneOverAinRiemann} as the black-hole spin $a$ varies~\footnote{The right half-plane corresponds to the conventional Riemann sheet, while the left half-plane corresponds to the unconventional sheet. Retrograde modes are not shown in Fig.~\ref{fig:QNM_l2_m2_frequencies}, since they lie on the conventional sheet in the left half-plane.}.
If one follows only the conventional sheet, namely the right half-plane in Fig.~\ref{fig:QNM_l2_m2_frequencies}, one observes the apparent discontinuity between negative and positive spins near the AS frequency.
Specifically, we see only a single $8_1$ mode for $a<0$, but both $8_0$ and $8_1$ modes for $a>0$; although these positive-spin branches head toward the AS frequency as $a\to+0$, they terminate at finite nonzero spin after hitting the branch cut.
This shows that the apparent discontinuity is not caused by the splitting of a single branch, but rather by the emergence onto the conventional sheet of a mode that was already present behind the branch cut on the unconventional sheet.
However, this global continuation alone does not explain why the trajectories take such a peculiar form, nor what kind of mode interaction occurs near the AS frequency.

To reveal the anatomy of the near-AS anomaly, we zoom in on the region around the negative AS frequency $M\omega=-2\mathrm{i}$ in the upper-left panel of Fig.~\ref{fig:qnm_and_B}.
As the spin approaches the Schwarzschild limit from the negative side, $a\to -0$, the $8_0$ mode on the unconventional sheet approaches the frequency identified in Ref.~\cite{Leung:2003eq}, whereas the $8_1$ mode remains on the conventional sheet and tends toward the negative AS frequency, corresponding to the eighth overtone in Leaver's table~\cite{Leaver:1985ax}. 
As the spin increases from $a=0$, the two branches approach each other but do not collide; instead, they exhibit a clear mutual repulsion.

This behavior closely parallels avoided-crossing structures previously found in Kerr QNM spectra at high spins~\cite{Motohashi:2024fwt}, but here it occurs in the immediate vicinity of the AS frequency and at an extremely small spin, $a/M\simeq1.5\times10^{-4}$. 
In the complex-frequency plane, the trajectories form the characteristic hyperbolic pattern of an avoided crossing. 
Thus, the apparent bifurcation seen in Fig.~\ref{fig:QNM_l2_m2_frequencies} is not a genuine splitting into disconnected branches, but the visible projection of an interaction between neighboring modes once the unconventional sheet is taken into account.

We further investigate the corresponding excitation factors~\cite{Leaver:1986gd,Sun:1988tz,Andersson:1995zk}, which quantify how efficiently individual modes are excited. For the Teukolsky variable, the excitation factor is defined by~\cite{Berti:2006wq,Zhang:2013ksa}
\begin{align}
    B_n^\mathrm{T} = \left. A^\mathrm{T}_\mathrm{out} / \bigl(2\omega {A^\mathrm{T}_\mathrm{in}}'\bigr)\right|_{\omega=\omega_n},
    \label{eq:excitationfactor}
\end{align}
where $A^\mathrm{T}_\mathrm{out}$ is the reflection amplitude and the prime denotes the derivative with respect to $\omega$. 
The excitation factors correspond to the residues associated with QNM poles and enter the gravitational-wave strain $h$ measured at a distance $r$ from the black hole as
\begin{align}
    h(t) \propto r^{-1}\sum_n \omega_n^{-2}B_n I_n e^{-\mathrm{i}\omega_n t},
\end{align}
where $I_n$ represents the source-dependent term (see Ref.~\cite{Kubota:2025hjk} for its explicit definition).
Recent high-precision calculations and independent cross-validation have substantially clarified Kerr excitation factors~\cite{Motohashi:2024fwt,Lo:2025njp,Kubota:2025hjk}; see also Ref.~\cite{DellaRocca:2025zbe}.

Extending these results, we present the excitation factors for the two eighth-overtone branches, $8_0$ and $8_1$.
The upper-right and lower panels of Fig.~\ref{fig:qnm_and_B} display the magnitudes of $B_n^\mathrm{T}$ and their trajectories in the complex plane, respectively. 
In the spin range where the two modes undergo an avoided crossing, the excitation factors are markedly enhanced and trace the lemniscate structure characteristic of resonant excitation~\cite{Motohashi:2024fwt}.
Thus, the near-AS avoided crossing is not merely a geometric rearrangement of QNM trajectories across Riemann sheets; it is accompanied by a resonant enhancement of the pole residues, revealing a hidden mode interaction behind the conventional branch cut~\footnote{See Supplemental Material B for details of the apparent singular behavior of the excitation factors near the pole skipping at the AS frequency.}.

\secl{Apparent Disappearance as Pole Skipping}
After clarifying the avoided crossing and resonant excitation behind the apparent bifurcation, we now turn to the other anomalous feature, namely, the apparent disappearance of the mode as it approaches the AS frequency. 
Unlike the bifurcation, which can already be inferred from the pole trajectories in Fig.~\ref{fig:qnm_and_B}, the disappearance cannot be understood from the pole structure alone. 
To explain it, one must also track the zeros of the reciprocal incident amplitude $1/A^\mathrm{T}_\mathrm{in}$.

For gravitational perturbations of Kerr black holes, these zeros occur precisely at the Matsubara-mode (MM) frequencies~\cite{Motohashi:2026mbn},
\begin{align}
    \omega_\mathrm{MM} = \mu_\mathrm{H} - 2\pi \mathrm{i} T_\mathrm{H} j, \qquad j=3,4,5,\dots,
    \label{eq:Matsubara}
\end{align}
where $\mu_\mathrm{H}=m\Omega_\mathrm{H}$ is the horizon chemical potential for the azimuthal mode, $\Omega_\mathrm{H}=a/(2Mr_+)$ is the horizon angular velocity, $T_\mathrm{H}=\kappa/(2\pi)$ is the Hawking temperature, $\kappa=\sqrt{M^2-a^2}/(2Mr_+)$ is the surface gravity, and $r_+=M+\sqrt{M^2-a^2}$ is the horizon radius.
The MM frequencies referred to here are the black-hole analogues of the Matsubara frequencies familiar from finite-temperature quantum statistical mechanics, reflecting the thermal nature of the black hole, with the imaginary part fixed by the Hawking temperature $T_\mathrm{H}$ and the real part set by the horizon chemical potential $\mu_\mathrm{H}$, in direct analogy with a grand-canonical ensemble at finite chemical potential.
Thus, the thermal frequency scale of the horizon appears directly in the zero structure of the Green-function building blocks.
Related roles of MMs have also been discussed recently in other black-hole and potential-scattering problems~\cite{Kuntz:2025gdq,Arnaudo:2025uos}. 
We use this Kerr-specific zero structure to explain why the Kerr QNM appears to disappear near the AS frequency.

In the upper-left panel of Fig.~\ref{fig:qnm_and_B}, the dashed curve represents a zero of the reciprocal incident amplitude identified numerically, in precise agreement with the Matsubara frequency with index $j=8$ in Eq.~\eqref{eq:Matsubara}.
This immediately explains why the $8_1$ mode is absent at the AS frequency in the incident-amplitude criterion.
In the Schwarzschild limit $a\to 0$, the $8_1$ pole exactly coincides with the Matsubara zero, so that near this point
\begin{align}
    \frac{1}{A^\mathrm{T}_\mathrm{in}}
    \propto
    \frac{\omega-\omega_\mathrm{MM}}{\omega-\omega_\mathrm{QNM}}.
\end{align}
The pole is therefore canceled by a coincident zero and becomes invisible in $1/A^\mathrm{T}_\mathrm{in}$.
This is precisely the phenomenon known as pole skipping~\cite{Grozdanov:2017ajz,Blake:2017ris,Blake:2018leo,Grozdanov:2018kkt}.

\begin{figure*}[]
   \begin{minipage}[]{0.5\textwidth}
        \includegraphics[width=\columnwidth]{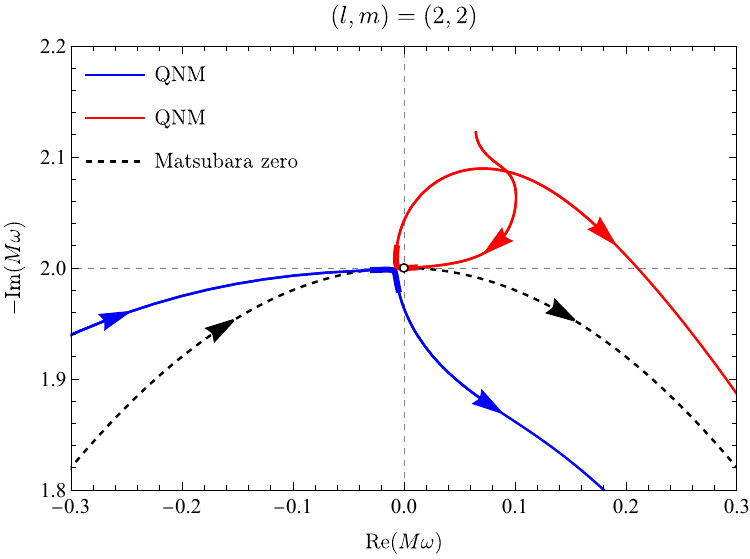}
    \end{minipage}
    \hspace{0.05\columnwidth}
    \begin{minipage}[]{0.37\textwidth}
        \includegraphics[width=\columnwidth]{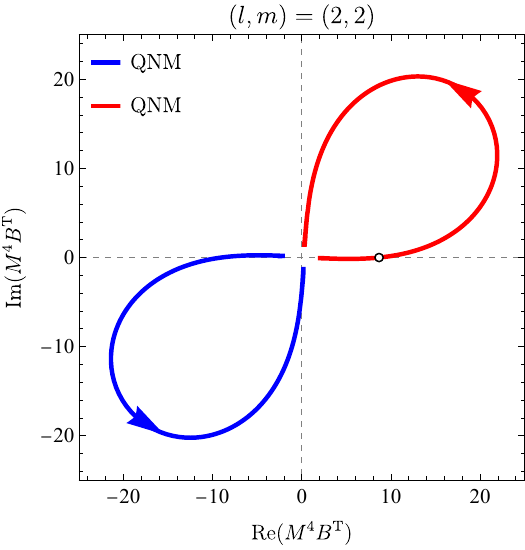}
    \end{minipage}
    \vspace{0.5cm}
    \\
   \begin{minipage}[]{0.5\textwidth}
        \includegraphics[width=1.0\columnwidth]{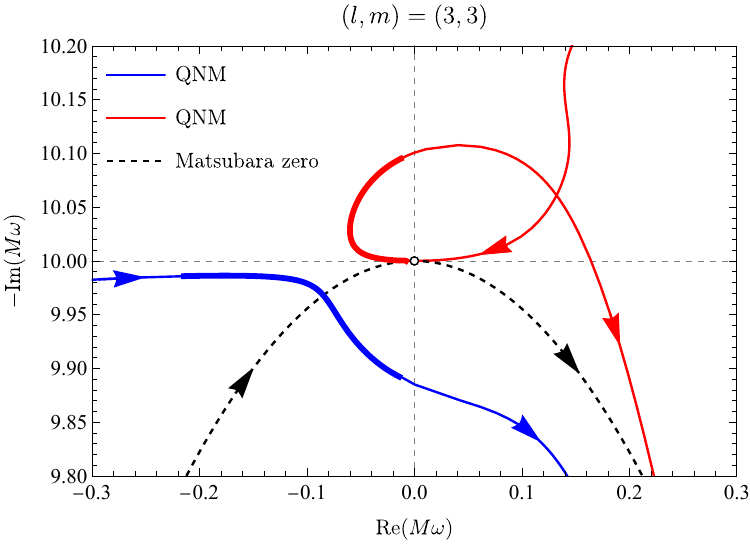}
    \end{minipage}
    \hspace{0.05\columnwidth}
    \begin{minipage}[]{0.37\textwidth}
        \includegraphics[width=1.0\columnwidth]{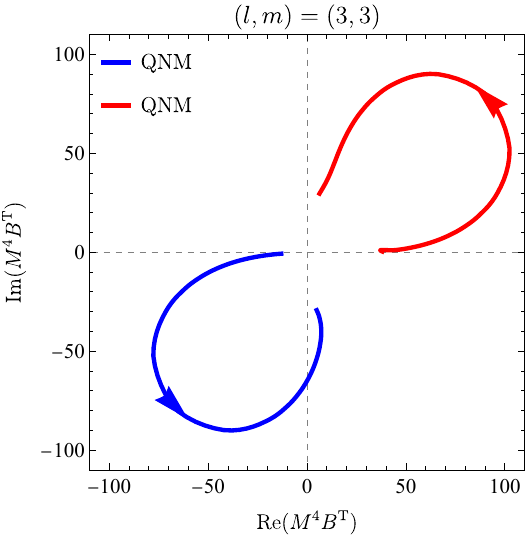}
    \end{minipage}
    \vspace{0.5cm}
    \\
    \begin{minipage}[]{0.5\textwidth}
        \includegraphics[width=1.0\columnwidth]{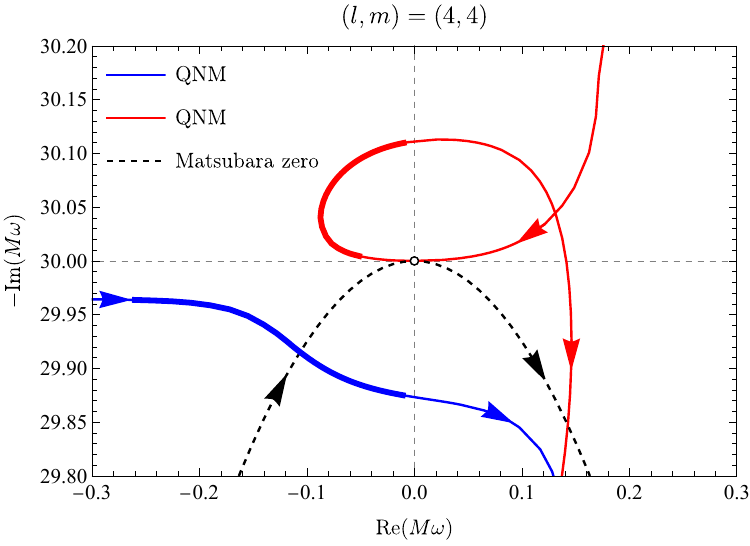}
    \end{minipage}  
    \hspace{0.05\columnwidth}
    \begin{minipage}[]{0.37\textwidth}
        \includegraphics[width=1.0\columnwidth]{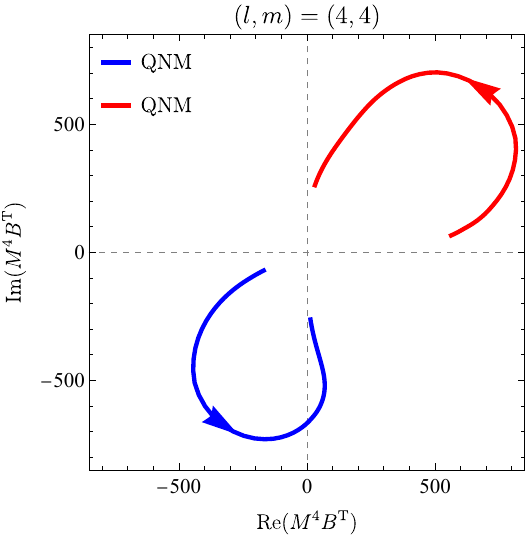}
    \end{minipage}
    \caption{\label{fig:qnm_228_qnm_228__44120}QNM poles (solid), Matsubara zeros (dashed) (left panels), and excitation factors (right panels) for $(l,m)=(2,2)$, $(3,3)$, and $(4,4)$. White circles mark pole-skipping points, arrows indicate increasing spin, and thick curves highlight the resonant-excitation regions, corresponding to the ranges $-2.9 \times 10^{-3} \leq a/M \leq 7.1 \times 10^{-3}$,  $0.5 \times 10^{-3} \leq a/M \leq 4.0 \times 10^{-3}$, and $0.72 \times 10^{-3} \leq a/M \leq 1.9 \times 10^{-3}$, respectively.}
\end{figure*}

This mechanism is not unique to the $(l,m)=(2,2)$ eighth overtone. At $a=0$, the MM frequencies satisfy $M\omega_\mathrm{MM}=-\mathrm{i}j/4$, while the AS frequencies are given by Eq.~\eqref{eq:ASfrequency}; these coincide when $j=(l-1)l(l+1)(l+2)/3$, which is an integer for every gravitational multipole $l\ge 2$; see Supplemental Material C.
More generally, AS frequencies are pole-skipping points for gravitational perturbations~\cite{Grozdanov:2023txs,Grozdanov:2025ner}.
Figure~\ref{fig:qnm_228_qnm_228__44120} shows that the same intertwined structure is also present near the AS frequencies for $(l,m)=(3,3)$ and $(4,4)$. 
In all three cases, the QNM coincides with the Matsubara zero in the nonrotating limit, while the avoided crossing becomes progressively milder for higher multipoles. 
These examples support the interpretation that the $(2,2)$ eighth overtone is the clearest representative of a broader class of Kerr modes near AS frequencies exhibiting the same pole-skipping, avoided-crossing, and resonant-excitation structure.

Thus, the Kerr mode approaching the AS frequency does not simply terminate; rather, the corresponding QNM pole is canceled by the Matsubara zero in the reciprocal incident amplitude.
This gives a precise criterion-dependent interpretation of whether the mode approaching the AS frequency should be regarded as a genuine QNM.
Under the usual incident-amplitude criterion, the coincidence of the pole and zero makes the mode invisible at the AS point~\footnote{The corresponding subtleties in alternative definitions, as well as the residual singularities in other Green-function building blocks, are discussed in the Supplemental Material D.}

\secl{\label{sec:conclusion}Conclusions and Discussion}
We have resolved the physical origin of a long-standing anomaly in Kerr quasinormal modes (QNMs) near algebraically special (AS) frequencies through a unified Green-function picture. 
The apparent bifurcation is not a splitting into disconnected branches, but an avoided crossing with a neighboring mode on an unconventional Riemann sheet. 
The same mode interaction produces resonant excitation, with the excitation factors exhibiting the characteristic enhancement and lemniscate structure expected near an avoided crossing.

The apparent disappearance of the mode has a different, but intertwined, origin: pole skipping. 
The mode therefore does not simply terminate; its disappearance in the standard criterion reflects a pole-zero collision fixed by the analytic structure of the reciprocal incident amplitude. 
Moreover, the AS frequencies are pole-skipping points for arbitrary gravitational multipoles, and our analysis shows that the same pole-skipping, avoided-crossing, and resonant-excitation structure persists beyond the representative $(2,2)$ case.
Since AS frequencies are tied to total-transmission behavior and occur only for gravitational perturbations, the anomaly exposes a genuinely gravitational feature of black-hole scattering rather than an accidental peculiarity of a particular overtone.

An important lesson is that black-hole spectroscopy cannot be understood from QNM pole trajectories alone. 
Zeros of Green-function building blocks, although much less explored in the ringdown literature, can control whether a QNM pole is visible or hidden in the corresponding Green-function contribution.
In the present case, these zeros occur at Matsubara frequencies set by the Hawking temperature and horizon chemical potential, suggesting that QNM spectra retain an analytic imprint of horizon thermality in gravitational scattering.

More broadly, our analysis recasts a long-standing relativistic puzzle as non-Hermitian pole dynamics in a black-hole setting. 
As the black-hole spin is varied, QNM poles move across Riemann sheets, undergo avoided crossings with resonant enhancement of their residues, and can be hidden by zeros of Green-function building blocks.
This places Kerr black-hole perturbations within a wider S-matrix perspective on open resonant systems, where poles and zeros move, collide, and exchange visibility across different Riemann sheets. 
It also establishes pole-zero tracking across Riemann sheets as a useful diagnostic for assessing black-hole QNM spectra, and connects black-hole perturbation theory with the broader language of pole dynamics in hydrodynamics, statistical physics, quantum optics, and other non-Hermitian systems.

\vspace{3mm}
\secl{Acknowledgments}
We thank Tetsuo Hyodo for useful discussion. 
For certain computations in this work, we used a modified version of the \texttt{Teukolsky} package from the \texttt{Black Hole Perturbation Toolkit}~\cite{BHPToolkit}. 
This work was supported by JSPS KAKENHI Grant No.~JP22K03639 and No.~JP26K21813 (H.M.).

\vspace{3mm}
\secl{Data availability}
The animations corresponding to the figures are openly available in Ref.~\cite{kubota_2026_18534214}.

\section*{Supplemental Material}

This Supplemental Material presents technical details and higher-multipole analyses supporting the claim in the main text that the intertwined pole-skipping, avoided-crossing, and resonant-excitation structure arises generically near algebraically special (AS) frequencies.

\appendix
\renewcommand{\appendixname}{}
\counterwithin{figure}{section}
\renewcommand{\thefigure}{\Alph{section}\arabic{figure}}

\section{Branch-cut rotation and Riemann-sheet structure\label{sec:numericalmethodology}}
Quasinormal modes (QNMs) satisfy purely ingoing boundary conditions at the horizon and purely outgoing boundary conditions at infinity. 
Standard techniques for determining QNM frequencies, such as Leaver's continued-fraction method~\cite{Leaver:1985ax}, encounter a branch cut along the negative imaginary axis. 
This conventional choice is inconvenient for investigating QNMs near AS frequencies.
To circumvent this issue, we rotate the branch cut by employing the Mano--Suzuki--Takasugi (MST) method~\cite{Mano:1996vt,Sasaki:2003xr}.

The in-mode solution of the homogeneous radial Teukolsky equation $R^\text{in}$ satisfies the boundary condition:
\begin{align}
    R^\text{in}(r)\to
    \left\{ \,
        \begin{aligned}
            &\Delta^2e^{-i(\omega-\mu_\mathrm{H})r_*} & \text{for } r\to r_+ , \\
            &r^3A^\text{T}_\mathrm{out}e^{i\omega r_*} + r^{-1}A^\text{T}_\mathrm{in}e^{-i\omega r_*} & \text{for } r\to \infty .
        \end{aligned}
    \right.\label{eq:Rinasympform}
\end{align}
QNMs are determined by the condition $A^\mathrm{T}_\mathrm{in}=0$. 
In the MST method, the incident amplitude can be expressed analytically
\begin{align}
A^\mathrm{T}_\mathrm{in} &= \left(K_\nu - i e^{-i\pi\nu}\frac{\sin(\pi(\nu-s+i\epsilon))}{\sin(\pi(\nu+s-i\epsilon))}\right) \notag\\
&\quad \times\omega^{-1} A_+^\nu e^{-i\epsilon\ln\epsilon - \frac{1-\kappa}{2}\epsilon},
\end{align}
with $\epsilon=2M\omega$. 
The definitions of the parameters $\nu$, $A_+$, and $K_\nu$ are given in Ref.~\cite{Sasaki:2003xr}.
The branch cut in this expression arises from the logarithmic term $\ln(2M\omega)$ in the exponent and from the noninteger power contained in $K_\nu$. 
To place the branch cut along an arbitrary ray $\arg z=\theta$, we redefine the complex logarithm and power as
\begin{align}
    \ln z &\coloneqq \ln|z| + \mathrm{i}\left[\theta + \mathrm{mod}(\arg z-\theta,2\pi)\right],\\
    z^a &\coloneqq |z|^a \exp\!\left\{\mathrm{i} a\left[\theta + \mathrm{mod}(\arg z-\theta,2\pi)\right]\right\}.
\end{align}
By varying $\theta$, we can evaluate the incident amplitude across multiple Riemann sheets. 

\begin{figure*}[htbp]
    \includegraphics[width=0.32\textwidth]{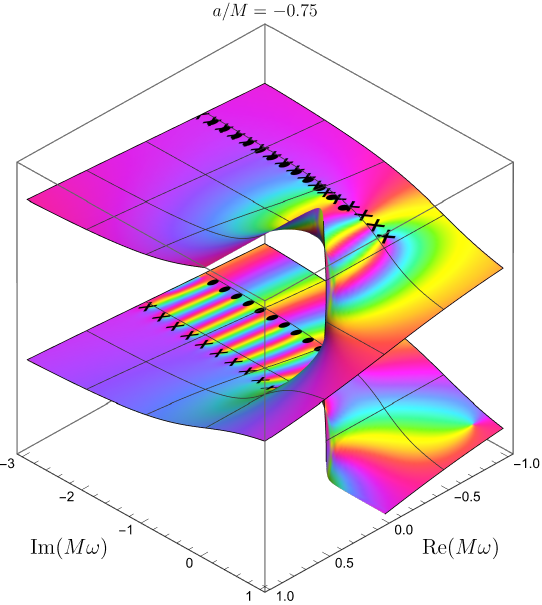}
    \includegraphics[width=0.32\textwidth]{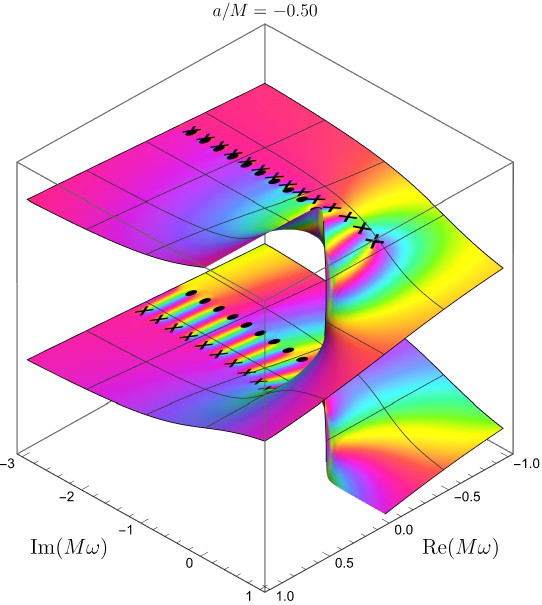}
    \includegraphics[width=0.32\textwidth]{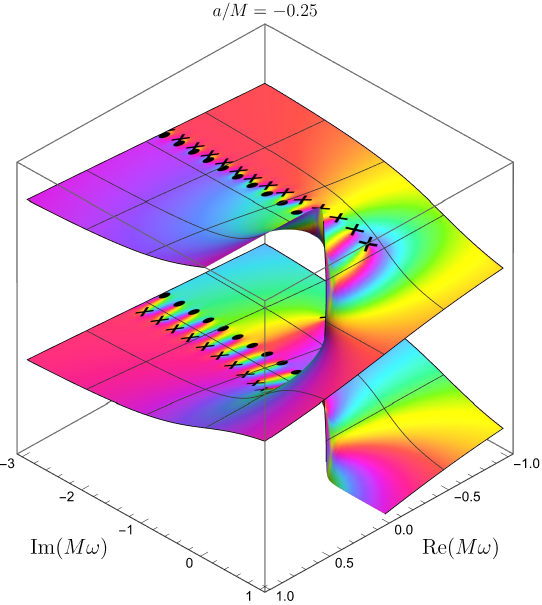}
    \includegraphics[width=0.32\textwidth]{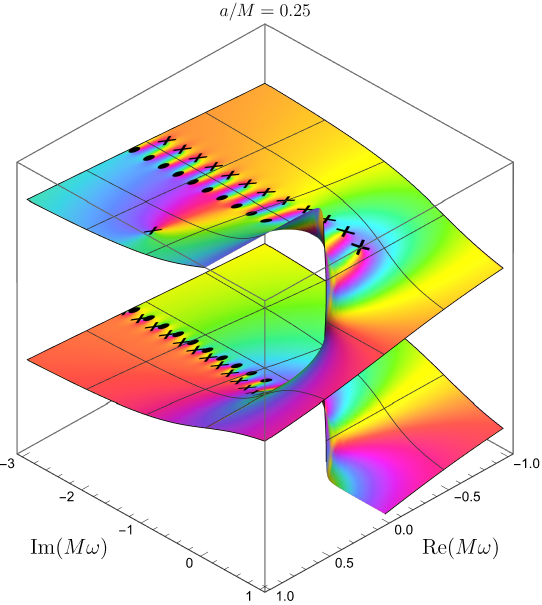}
    \includegraphics[width=0.32\textwidth]{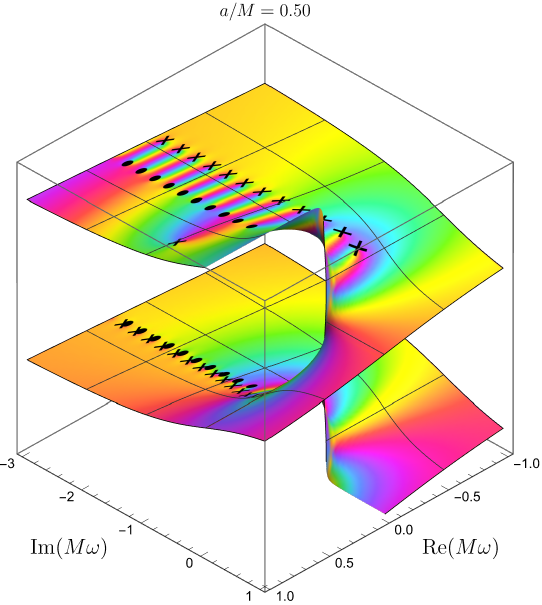}
    \includegraphics[width=0.32\textwidth]{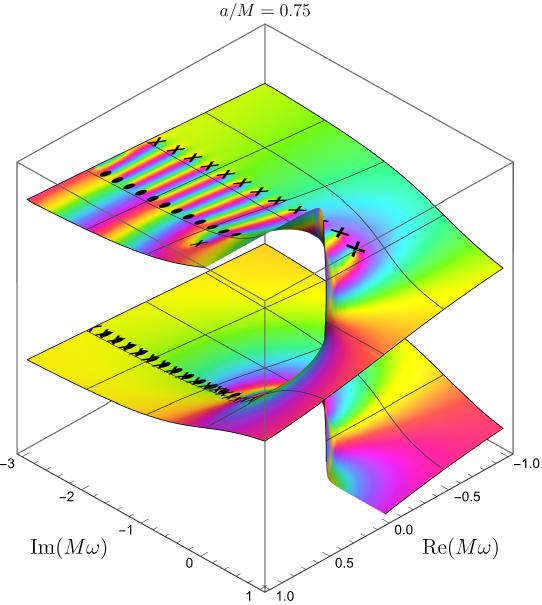}
    \caption{\label{fig:OneOverAinRiemannKerr}Snapshots of the Riemann-sheet structure of $1/A^\mathrm{T}_\mathrm{in}$ for Kerr black holes at representative spin values. Markers and colors follow Fig.~1; animations are available in Ref.~\cite{kubota_2026_18534214}.}
\end{figure*}

Figure~\ref{fig:OneOverAinRiemannKerr} shows snapshots of the pole-and-zero structure of $1/A^\mathrm{T}_\mathrm{in}$ for Kerr black holes at representative spin values. 
Animations with finer spin increments are available in Ref.~\cite{kubota_2026_18534214}.

\section{Apparent divergence of excitation factor\label{sec:apparentdivergence}}

\begin{figure*}[]
   \begin{minipage}[]{\columnwidth}
        \includegraphics[width=1.0\columnwidth]{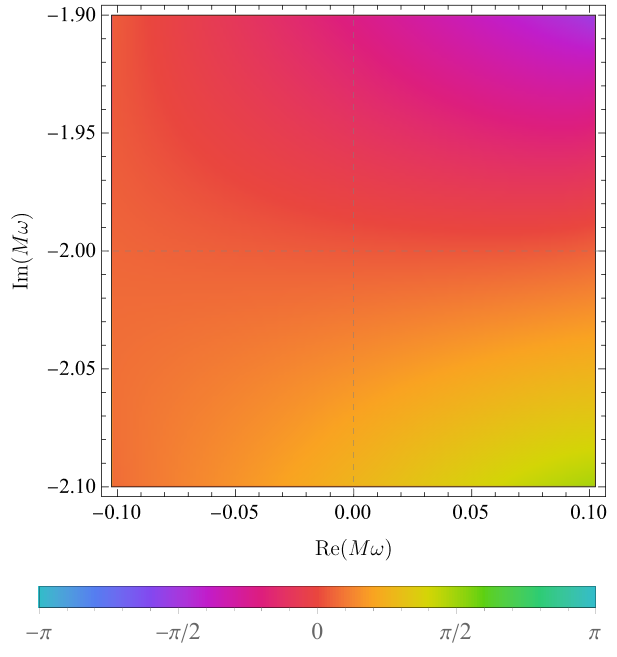}
    \end{minipage}
    \begin{minipage}[]{\columnwidth}
        \includegraphics[width=1.0\columnwidth]{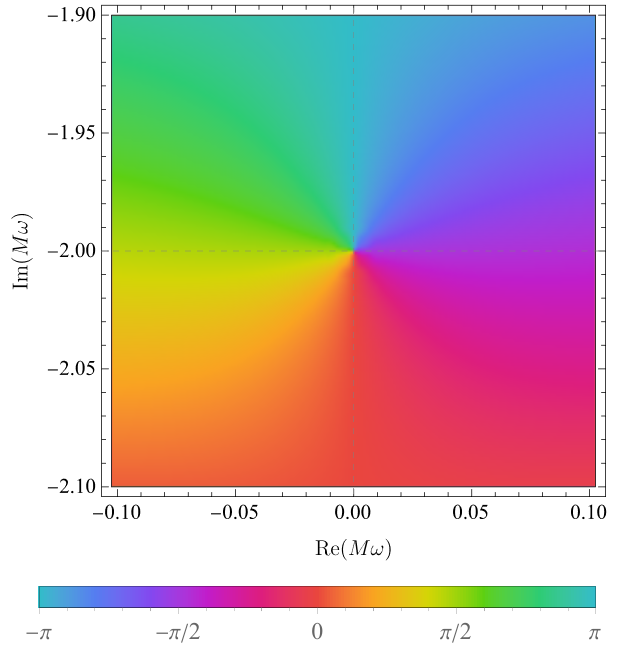}
    \end{minipage}
    \caption{\label{fig:Ain_prime_contourplot_a0andAout_contourplot_a0}Phase of ${A^\mathrm{T}_\mathrm{in}}'$ (left) and $A^\mathrm{T}_\mathrm{out}$ (right) for $(l,m)=(2,2)$ at $a=0$.
    }
\end{figure*} 

We next examine the behavior of the Teukolsky excitation factor 
$B_n^\mathrm{T} = \left. A^\mathrm{T}_\mathrm{out} / \bigl(2\omega {A^\mathrm{T}_\mathrm{in}}'\bigr)\right|_{\omega=\omega_n}$
at the pole-skipping point. 
For the $(l,m)=(2,2)$ mode shown in Fig.~\ref{fig:Ain_prime_contourplot_a0andAout_contourplot_a0}, ${A^\mathrm{T}_\mathrm{in}}'$ remains finite and nonzero at $a=0$, whereas $A^\mathrm{T}_\mathrm{out}$ has a pole at the AS frequency. 
Consequently, the excitation factor $B_n^\mathrm{T}$ diverges formally at the pole-skipping point.
We have verified that the same qualitative behavior also occurs for the higher-multipole cases $(l,m)=(3,3)$ and $(4,4)$.

This apparent divergence does not imply a divergence of the physical waveform. 
The $n$th QNM amplitude in the gravitational-wave strain is proportional to $\omega_n^{-2}B_n I_n$.
Here, the source-dependent term $I_n$ contains a compensating factor $1/A^\mathrm{T}_\mathrm{out}$, which cancels the singularity in $B_n^\mathrm{T} \propto A^\mathrm{T}_\mathrm{out}$.
Thus, the gravitational-wave amplitude remains finite even in the limit toward the pole-skipping point.

The same conclusion is obtained in the Sasaki--Nakamura (SN) or Regge--Wheeler (RW) description for general multipoles. 
The Teukolsky and SN/RW excitation factors are related via~\cite{Zhang:2013ksa}
\begin{align}
    B_{n}^\mathrm{SN} = \frac{c_0}{(2\omega)^4}B_{n}^\mathrm{T}
\end{align}
with 
\begin{align} \label{eq:c0Kerr}
c_0=\lambda(\lambda+2)-12\mathrm{i}M\omega-12a\omega(a\omega-m).
\end{align}
In the Schwarzschild limit, since $\lambda=l(l+1)-s(s+1)$ with $s=-2$, this reduces to
\begin{align}
\label{eq:c0Sch} c_0 = -12\mathrm{i} \left( M\omega+\mathrm{i} \frac{(l-1)l(l+1)(l+2)}{12} \right), 
\end{align}
which vanishes precisely at the negative AS frequency.
The zero of $c_0$ therefore cancels the pole in $B_n^\mathrm{T}$, and the SN/RW excitation factor remains finite. 
Hence pole skipping at the AS frequency does not produce a divergent physical response.

\section{Anomalous behavior for higher multipoles\label{sec:anomalousbehaviorforhigherovertone}}

\begin{figure*}[t]
    \includegraphics[width=1.0\textwidth]{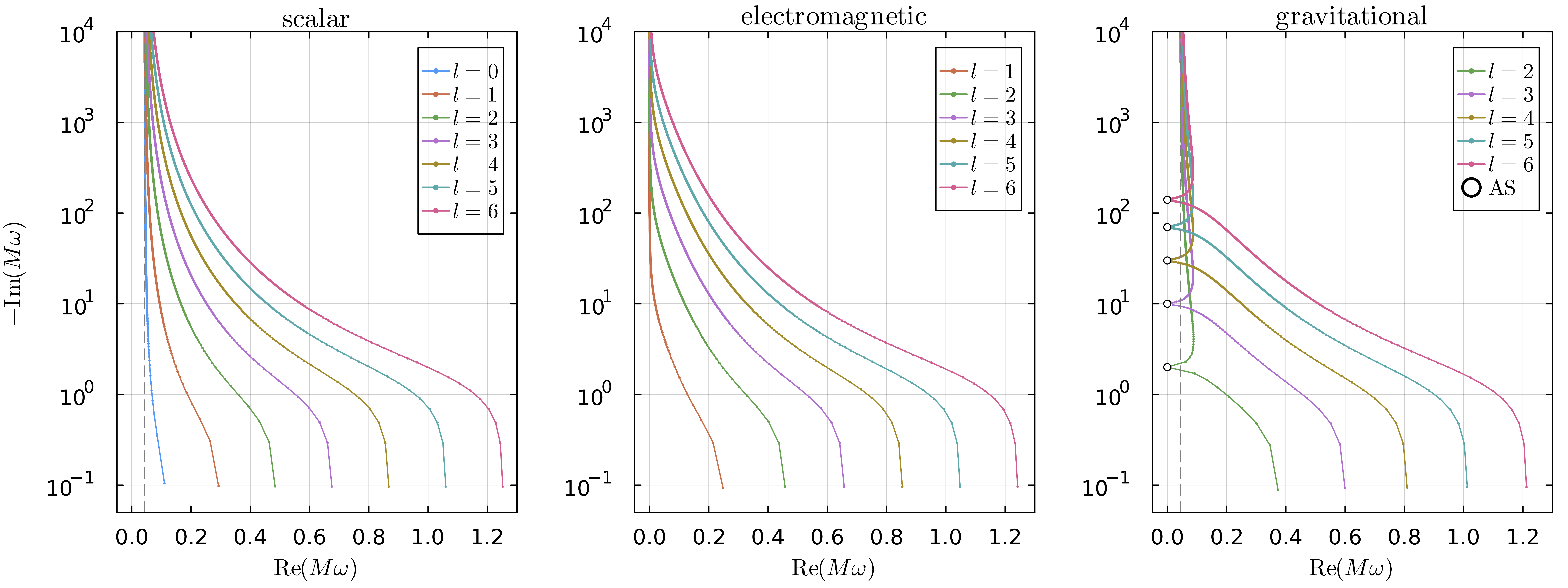}
    \caption{\label{fig:QNMFrequenciesInSchwarzschildBH}Schwarzschild QNM spectra for scalar (left), electromagnetic (middle), and gravitational (right) perturbations. White circles mark AS frequencies, and the grey dashed line marks $\ln(3)/(8\pi)$. AS pole-skipping points appear only for gravitational perturbations.}
\end{figure*}

The near-AS anomalous features identified in the main text—pole skipping, avoided crossing, and resonant excitation—are not unique to the $(l,m)=(2,2)$ multipole: the same combined behavior is shown in the main text for the $(3,3)$ and $(4,4)$ multipoles. The pole-skipping structure itself persists for even higher multipoles.
Figure~\ref{fig:QNMFrequenciesInSchwarzschildBH} shows the Schwarzschild QNM spectra for scalar, electromagnetic, and gravitational perturbations together with the AS frequencies, where we compute high-overtone frequencies up to $\Im(M\omega)=-10^4$ based on the procedure developed by Nollert~\cite{Nollert:1993zz}. 
Only in the gravitational case does the QNM trajectory approach the AS frequency before tending to its well-known asymptotic real part~\cite{Hod:1998vk,Andersson:2003fh}. 
Since AS frequencies coincide with Matsubara zeros, they are in general pole-skipping points~\cite{Grozdanov:2023txs}, a mechanism specific to gravitational perturbations.
Numerically, we confirm this pole-skipping structure at least up to $l=6$.

\section{QNM criteria at the AS frequency\label{sec:AmbiguityinQNMDefinition}}
We now revisit whether the AS mode should be regarded as a genuine QNM. 
Here we examine three notions of QNM behavior at the AS point—defined through the incident amplitude, the reflection coefficient, and the scattering matrix—taking pole skipping into account.
Below, we denote by $\omega_\mathrm{AS}$ the Schwarzschild AS frequency on the negative imaginary axis,
\begin{align}
    M\omega_\mathrm{AS} = - \mathrm{i} \frac{(l-1)l(l+1)(l+2)}{12}.
\end{align}

Before comparing these criteria, we emphasize a useful distinction. 
The Matsubara singularities do not define poles or zeros of the full Green's function in a representation-independent sense. 
In the Teukolsky fixed-sector connection problem, they appear in individual Green-function building blocks, and the explicit Matsubara factors normally cancel in the combinations entering decomposed Green-function contributions~\cite{Motohashi:2026mbn}. 
Schematically, a Teukolsky Green-function contribution contains a product of a homogeneous solution and an inverse incident amplitude, $G\sim R^\mathrm{in}R^\mathrm{up}/A^\mathrm{T}_\mathrm{in}$. 
Away from the AS point, both $R^\mathrm{in}$ and $A^\mathrm{T}_\mathrm{in}$ contain the same Matsubara pole factor, so this factor cancels in the assembled contribution~\cite{Motohashi:2026mbn}.
At the AS point, however, $A^\mathrm{T}_\mathrm{in}$ also has a QNM zero at the same frequency, so the Matsubara pole factor in $A^\mathrm{T}_\mathrm{in}$ is canceled within $A^\mathrm{T}_\mathrm{in}$ itself. 
As a result, the QNM pole is hidden in $1/A^\mathrm{T}_\mathrm{in}$, while the Matsubara pole of $R^\mathrm{in}$ can remain in the Teukolsky fixed-sector Green-function representation. 
Thus, pole skipping here should be understood as a statement about the reciprocal incident amplitude $1/A^\mathrm{T}_\mathrm{in}$, not as a statement that the total Green's function has no singular contribution at the AS frequency.

First, under the standard definition, a QNM frequency is a zero of the incident amplitude $A^\mathrm{T}_\mathrm{in}$. 
As discussed in the main text, a QNM pole approaching the AS frequency disappears because a coincident Matsubara-mode zero yields
\begin{align}
    \frac{1}{A^\mathrm{T}_\mathrm{in}} \propto \frac{\omega - \omega_\mathrm{MM}}{\omega - \omega_\mathrm{QNM}}.
\end{align}
Both $\omega_\mathrm{QNM}$ and $\omega_\mathrm{MM}$ approach $\omega_\mathrm{AS}$ as $a\to 0$, so the pole is canceled by the zero and becomes invisible in $1/A^\mathrm{T}_\mathrm{in}$.
For the SN variable, which reduces to the RW variable in the Schwarzschild limit, the incident amplitude is related to the Teukolsky amplitude by
\begin{align} \label{eq:SNinT}
    A^\mathrm{SN}_\mathrm{in} = -(2\omega)^2 A^\mathrm{T}_\mathrm{in}.
\end{align}
Since the corresponding SN/RW incident amplitude is therefore regular at the AS frequency, the AS mode is not classified as a QNM in this standard incident-amplitude sense.

Second, from Eq.~\eqref{eq:Rinasympform}, one may instead define a QNM as a pole of the reflection coefficient $A^\mathrm{T}_\mathrm{out}/A^\mathrm{T}_\mathrm{in}$. 
Under this definition, the AS mode is QNM-like for the Teukolsky variable because $A^\mathrm{T}_\mathrm{out}$ itself has a pole,
\begin{align}
    \frac{A^\mathrm{T}_\mathrm{out}}{A^\mathrm{T}_\mathrm{in}} \propto \frac{1}{\omega - \omega_\mathrm{MM}}\frac{\omega - \omega_\mathrm{MM}}{\omega - \omega_\mathrm{QNM}}.
\end{align}
For the SN/RW variable, however, the corresponding quantity remains finite. Using the relation~\cite{Sasaki:2003xr}
\begin{align}
    A^\mathrm{SN}_\mathrm{out} = -\frac{c_0}{4\omega^2} A^\mathrm{T}_\mathrm{out},
\end{align}
and Eq.~\eqref{eq:SNinT}, we obtain
\begin{align}
    \frac{A^\mathrm{SN}_\mathrm{out}}{A^\mathrm{SN}_\mathrm{in}} \propto \frac{\omega - \omega_{c0}}{\omega - \omega_\mathrm{MM}}\frac{\omega - \omega_\mathrm{MM}}{\omega - \omega_\mathrm{QNM}}.
\end{align}
Here, $\omega_{c0}$ is the frequency at which $c_0(\omega)=0$, which approaches the AS frequency in the Schwarzschild limit, as in Eq.~\eqref{eq:c0Sch}.
Therefore, in the Schwarzschild limit, $c_0$ cancels the pole in $A^\mathrm{T}_\mathrm{out}$ for general multipoles. 
This is consistent with the classic result that the AS frequency is absent from the RW spectrum~\cite{MaassenvandenBrink:2000iwh}.

Finally, the same conclusion is reflected in the scattering matrix. Following Refs.~\cite{Matzner1978,Futterman:1988ni,Dolan:2008kf,Leite:2017zyb,Markovic:2025kvr}, the scattering matrix $\mathcal{S}$ can be written as
\begin{align}\label{eq:S}
    \mathcal{S} =e^{2\mathrm{i}\delta} = (-1)^{l+1}\frac{\mathrm{Re}(C)+12\mathrm{i}M\omega P}{(2\omega)^4}\frac{A^\mathrm{T}_\mathrm{out}}{A^\mathrm{T}_\mathrm{in}},
\end{align}
where $P=\pm1$ denotes the parity, $\delta$ is the phase shift, and 
\begin{align}
C &= [\{(\lambda+2)^2 + 4a\omega-4a^2\omega^2\}(\lambda^2+36am\omega-36a^2\omega^2) \notag\\
&\quad +(2\lambda+3)(96a^2\omega^2-48am\omega)-(12a\omega)^2]^{1/2} \notag\\
&\quad + 12\mathrm{i}M\omega,
\end{align}
is the Teukolsky--Starobinsky constant.
In the Schwarzschild limit, the numerator of Eq.~\eqref{eq:S} behaves as
\begin{align}
    \mathrm{Re}(C)+12\mathrm{i}M\omega P \propto M\omega + \mathrm{i}P\,\frac{(l-1)l(l+1)(l+2)}{12},
\end{align}
with the sign chosen so as to reproduce the standard RW/Zerilli distinction at the AS frequency.
This implies that, at the negative AS frequency, the numerator vanishes only for even parity ($P=1$), whereas it remains finite for odd parity ($P=-1$).
Therefore the scattering matrix has a pole at the AS frequency only for odd parity, whereas for even parity that pole is canceled explicitly.
Again, this reproduces the established distinction between the RW (odd-parity) and Zerilli (even-parity) descriptions at the AS frequency~\cite{MaassenvandenBrink:2000iwh}.

\bibliography{references}
\end{document}